\documentclass[]{spie}  
\usepackage[]{graphicx}

\title{The Infrared Imaging Spectrograph (IRIS) for TMT: The Science Case} 

\author{Elizabeth J. Barton\supit{a}, James E. Larkin\supit{b}, Anna M. Moore\supit{c}, 
Shelley A. Wright\supit{d}, David Crampton\supit{e}, Luc Simard\supit{e}, Bruce Macintosh\supit{f},
Patrick C\^{o}t\'{e}\supit{e}, Aaron J. Barth\supit{a}, Andrea M. Ghez\supit{b}, Jessica R. Lu\supit{g},
T. J. Davidge\supit{e},David R. Law\supit{b}, and the IRIS Science Team
\skiplinehalf
\supit{a}Center for Cosmology, University of California, Irvine, USA; \\
\supit{b}Division of Astronomy \& Astrophysics, University of California, Los Angeles, USA; \\
\supit{c}Caltech Optical Observatories, California Institute of Technology, USA; \\
\supit{d}Department of Astronomy, University of California, Berkeley, USA; \\
\supit{e}Herzberg Institute of Astrophysics, National Research Council Canada\\
\supit{f}Lawrence Livermore National Lab., Livermore, CA, USA\\
\supit{g}California Institute of Technology, Pasadena, CA, USA\\
}

\authorinfo{Further author information: (Send correspondence to E.J.B.)\\
E.J.B.: E-mail: ebarton@uci.edu.edu, Telephone: 1 949 824 7178\\  
J.E.L.: E-mail: larkin@astro.ucla.edu, Telephone: 1 310 825 9470\\
A.M.M.: E-mail: amoore@astro.caltech.edu, Telephone: 1 626 395 8918\\
S.A.W.: E-mail: saw@astro.berkeley.edu, Telephone: 1 310 980 7060\\
D.C.: E-mail: David.Crampton@nrc-cnrc.gc.ca, Telephone: 1 250 363 0010\\
L.S.: E-mail: Luc.Simard@nrc-cnrc.gc.ca, Telephone: 1 250 363 8315\\
B.M.: E-mail: macintosh1@llnl.gov, Telephone: 1 925 423 8129\\
P.C.: E-mail: Patrick.Cote@nrc-cnrc.gc.ca, Telephone: 1 250 363 8133\\
A.J.B.: E-mail: barth@uci.edu, Telephone: 1 949 824 3013\\
A.M.G.: E-mail: ghez@astro.ucla.edu, Telephone: 1 310 206 0420\\
J.R.L.: E-mail:  jlu@astro.caltech.edu, Telephone 1 626 395 3693\\
T.J.D.: E-mail: Tim.Davidge@nrc-cnrc.gc.ca, Telephone: 1 250 363 0047\\
D.R.L.: E-mail: drlaw@astro.ucla.edu, Telephone: 1 310 794 9466\\
}

\newcommand{\farcs}{\mbox{\ensuremath{.\!\!^{\prime\prime}}}}

\begin{document} 
  \maketitle 

\begin{abstract}
The InfraRed Imaging Spectrograph (IRIS) is a first-light instrument
being designed for the Thirty Meter Telescope (TMT). IRIS is a
combination of an imager that will cover a 16$^{\farcs}$4 field
of view at the diffraction limit of TMT (4 mas sampling), and an
integral field unit spectrograph that will sample objects at 4-50 mas
scales. IRIS will open up new areas of observational parameter space,
allowing major progress in diverse fields of astronomy.  We present
the science case and resulting requirements for the performance of
IRIS.  Ultimately, the spectrograph will enable very well-resolved and
sensitive studies of the kinematics and internal chemical abundances
of high-redshift galaxies, shedding light on many scenarios for the
evolution of galaxies at early times.  With unprecedented imaging and
spectroscopy of exoplanets, IRIS will allow detailed exploration of a
range of planetary systems that are inaccessible with current
technology.  By revealing details about resolved stellar populations
in nearby galaxies, it will directly probe the formation of systems
like our own Milky Way.  Because it will be possible to directly
characterize the stellar initial mass function in many environments
and in galaxies outside of the the Milky Way, IRIS will enable a
greater understanding of whether stars form differently in diverse
conditions.  IRIS will reveal detailed kinematics in the centers of
low-mass galaxies, allowing a test of black hole formation scenarios.
Finally, it will revolutionize the characterization of reionization
and the first galaxies to form in the universe.
\end{abstract}



\section{INTRODUCTION}
\label{sec:intro}  

As a premier first-light instrument for the Thirty Meter Telescope,
IRIS\cite{Larkin2010} will open huge windows of astronomical discovery
space \cite{Wright2010b}, paving the way for a new era of exploration
with extremely large telescopes.  Operational in the near-infrared
(0.8-2.4 $\mu m$), IRIS will offer diffraction-limited imaging and
integral-field spectroscopy behind the TMT facility Narrow-Field
Infrared Adaptive Optics system (NFIRAOS)\cite{Herriot2010}.  With 4
spatial scales for spectroscopic sampling (4, 9, 25, and 50 mas), its
versatility will allow the efficient study of both unresolved and
diffuse objects.  Here, we describe a subsample of its many
breakthrough scientific capabilities.

\section{Imaging and spectroscopy of exoplanets} \label{sec:exoplanets}

The direct imaging detection of extrasolar planets, and their
characterization via spectroscopy, remain frontier topics in
astrophysical research.  In the exciting hunt for an understanding of
planets outside of our own Solar system, IRIS on TMT will combine
extreme sensitivity and spectroscopic resolution in the range of $R
\sim 1000-8000$ to complement special purpose planet hunting
instrumentation.  IRIS will also bridge the gap until high-contrast
instruments such as the Planet Formation Imager \cite{Macintosh2006a} come online for TMT.

\begin{figure}
   \begin{center}
   \begin{tabular}{c}
   \includegraphics[height=7cm]{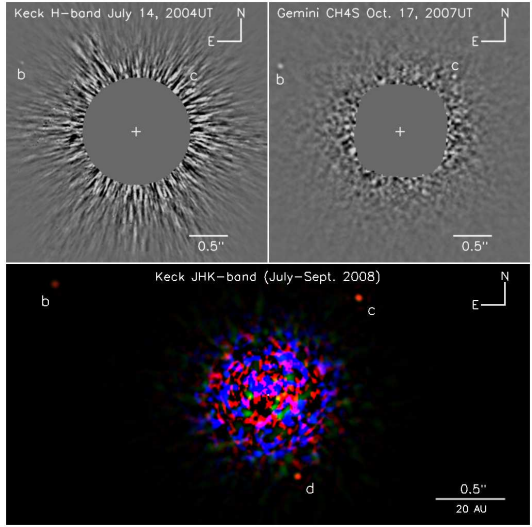}
   \end{tabular}
   \end{center}
   \caption[example] 
   { \label{fig:HR8799} 
HR8799bcd discovery images from \cite{Marois2008}, including 
({\it Upper left}) a Keck image, ({\it Upper right}) a Gemini image, and
({\it Bottom}) a color image from the Keck data showing three planets around the parent
star.  The combination of TMT and IRIS will excel at detecting relatively 
low-contrast planets that are as close as $\sim0\farcs1$ to their parent stars.}
   \end{figure}

IRIS will excel at higher-resolution ($R \geq 200-1000$)
spectroscopic study of known self-luminous extrasolar planets at large
separations, where scattered light is not the limiting factor.
Examples include the companions of HR8799 \cite{Marois2008} shown in
Fig~\ref{fig:HR8799}.  Intense spectroscopic study of these and other
planetary companions will yield unprecedented information about their
gravity, composition, and atmospheres.  Thus, this science case drives
requirements on the sensitivity of IRIS, as well as the point-spread
function and the dynamic range.

Also at the forefront of extrasolar planetary research, the angular
resolution afforded by the thirty meter primary aperture of TMT will
allow IRIS to excel over other planned instrumentation at discovering
bright planets that are {\it very} close to their parent stars, $\leq
0\farcs1$, which corresponds to 10-15 AU in star-forming regions and
$\sim$5 AU in young stellar associations.  These objects would include
forming protoplanets in, e.g., Taurus and Ophiuchus, which are often
too dim and red for higher-contrast AO systems like the Gemini Planet
Imager \cite{Macintosh2006b}.

\section{Black holes}

The mysterious black holes predicted by Einstein's theory of General
Relativity (GR) provide remarkable laboratories for the study of
physics in extreme environments.  Progress in this field drives
instrument requirements on the combination of angular resolution,
which allows gravitational tests much closer to the black holes
themselves and will make use of the diffraction-limited capabilities
of the instrument, and sensitivity, which is required to probe a
larger and/or more abundant population of ``test masses'' in order to
understand the black holes' effects.  In the case of the Galactic
Center, the need to trace the orbits of stars also drives the strict
astrometric requirements on IRIS (50 $\mu$as).  Thus IRIS on TMT will
be an ideal instrument with which to study these elusive bodies.

\subsection{Fundamental physics from the black hole at the center of the Galaxy} \label{sec:GalacticCenter}

\begin{figure}
   \begin{center}
   \begin{tabular}{c}
   \includegraphics[width=6.5in]{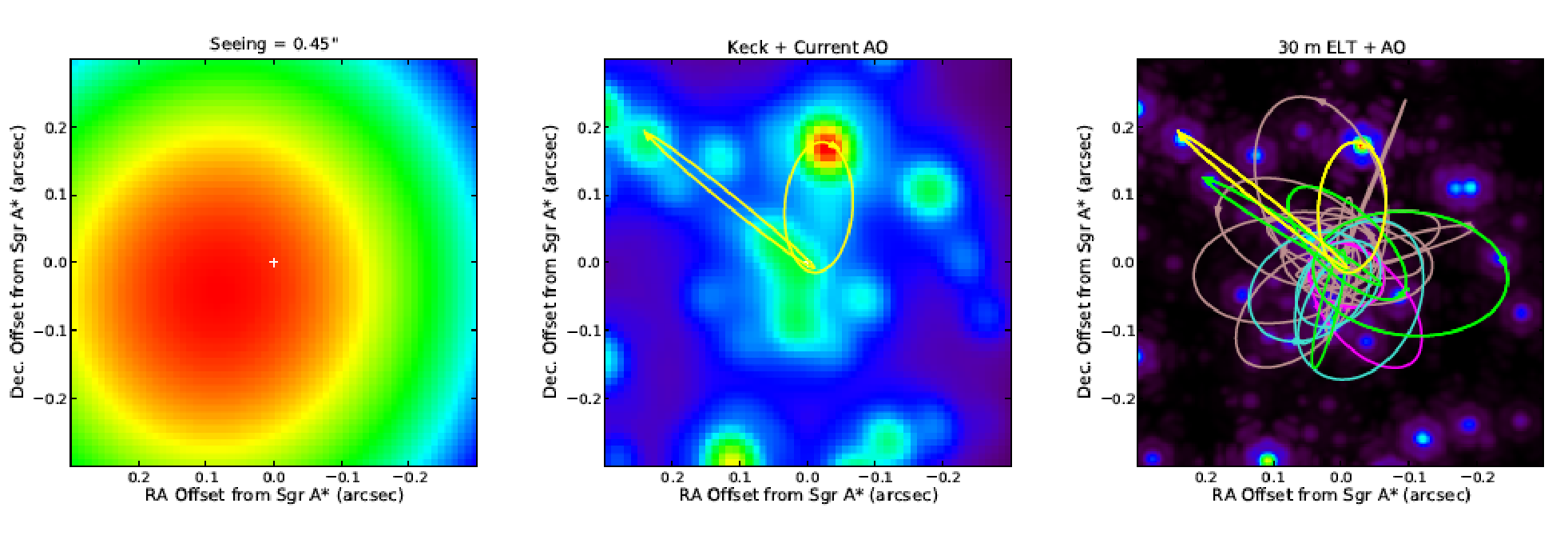}
   \end{tabular}
   \end{center}
   \caption[example] 
   { \label{fig:GalacticCenter} 

Probing the Galactic Center with TMT, courtesy of A. Ghez,
N. Weinberg, M. Morris, and J. Lu.  The figure shows a seeing-limited
observation of the $0\farcs6$ region of the Galactic Center ({\it
Left}), an image with the current Keck adaptive optics capabilities
({\it Middle}), and a simulation of the Galactic Center as observed by
TMT ({\it Right}).  The IRIS instrument on TMT will reveal the orbits
of many fainter and closer probes of the Galactic Center potential
well..}

   \end{figure}

For the first time, IRIS will extend our knowledge of the spacetime
topology around a supermassive black hole to well within $\sim$1000
Schwarzchild radii (see Fig.~\ref{fig:GalacticCenter}).  The immediate
surroundings of the black hole at the center of our Galaxy will allow
fundamental tests of ``medium field'' relativity, including the
astrometric signal of prograde GR precession \cite{Weinberg2005}, 
the influence on radial velocity measurements from the
special relativistic transverse Doppler shift, and the general
relativistic gravitational redshift \cite{Zucker2006}.

Because the detailed orbits of stars in the vicinity of the central
black hole will be affected by the presence or absence of dark matter,
the orbits of stars within 0.01 pc of the Galactic Center will
significantly constrain the dark matter models on which galaxy
formation is built.  With a decade of astrometric measurements using
IRIS on TMT, this orbital influence would be readily detectable,
standing ready to confirm theoretical predictions of the amount of
Galactic dark matter within the very center of the Galaxy.  Finally,
at its astrometric requirement ($\sim$50-100 $\mu$as), IRIS on TMT
will provide a very precise measurement of the distance to the
Galactic Center, R$_0$. A measure within 1\% would enable an accurate
determination of the size and shape of the Milky Way's dark matter
halo on 100-kpc scales \cite{Olling2000}.

\subsection{Black holes in other galaxies} \label{sec:BHOtherGalaxies}

\begin{figure}
   \begin{center}
   \begin{tabular}{c}
   \includegraphics[height=3in]{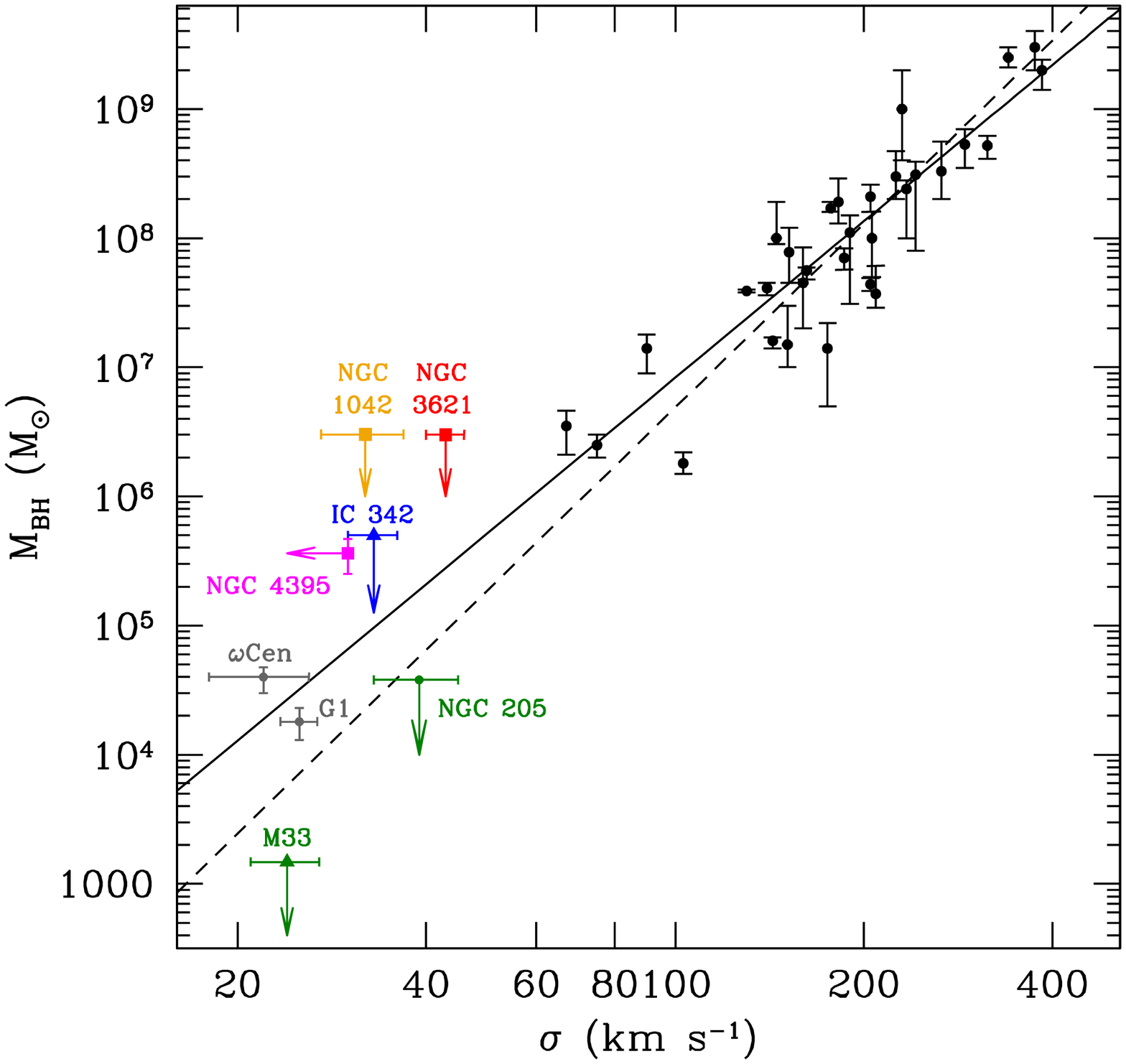}
   \includegraphics[height=3in]{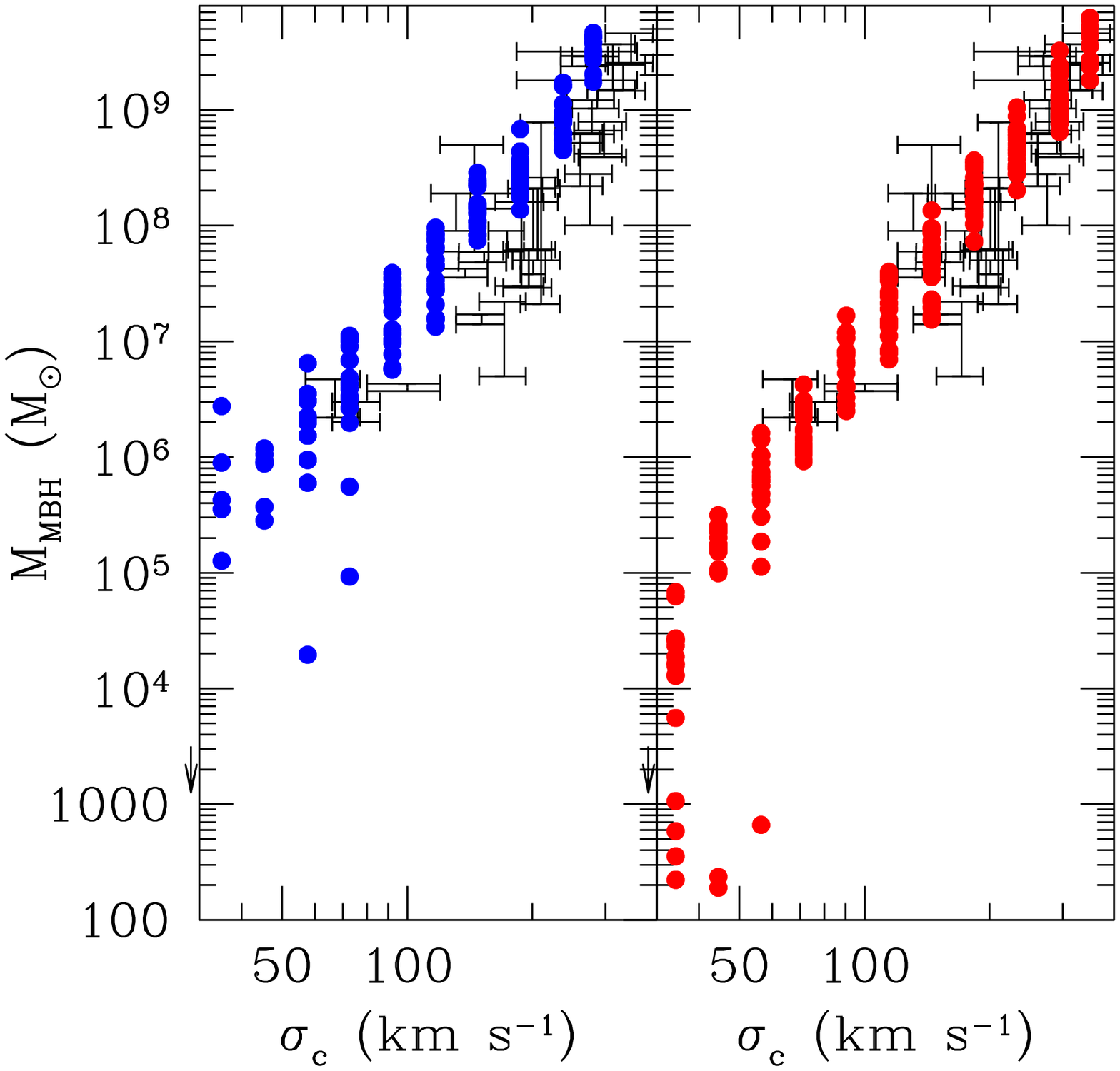}
   \end{tabular}
   \end{center}
   \caption[example] 
   { \label{fig:BlackHoles} 

({\it Left}) The black hole mass -- bulge velocity dispersion relation
from Barth et al. (2009a), based on new work and a compilation from
the literature.  The relation demonstrates the deep connections
between the formation of bulges and spheroids and the formation of
central supermassive black holes in galaxies.  ({\it Right}) Models of
black hole masses from \citenum{Barth2009}, using two formation
scenarios adapted from \citenum{Volonteri2008}.  The model on the left
has 10$^5$ solar mass seed black holes, while the model on the right
starts with 100-solar-mass black holes originating from Population III
stars.  IRIS will measure the masses of low-mass black holes in the
nearby universe, distinguishing between these scenarios.  }

   \end{figure}

Since the discovery that black holes reside in the centers of most, if
not all, massive galaxies, their role in galaxy formation and
evolution has been at the forefront of astronomical research.  Direct
measurements of their masses require both spatially well-resolved
spectroscopy at high spectral resolution ($R \sim 4000-8000$).  For
the first time, the sensitivity afforded by IRIS on TMT will allow
these measurements to extend to more distant galaxies and to fainter,
lower mass galaxies.  Thus, it will extend our knowledge of both the
high-mass and low-mass ends of the relationship between black hole
masses and bulge velocity dispersion, $\sigma$ (Fig.~\ref{fig:BlackHoles}), which is
a fundamental probe of the relationship between black hole growth and galaxy
formation.

Low-mass black holes and galactic nuclei are of particular interest
for black hole formation scenarios.  Recent studies with the Hubble
Space Telescope (HST) have shown that the vast majority of
low-luminosity galaxies contain compact stellar nuclei (sometimes
referred to as nuclear star clusters) whose properties are also
connected to those of their hosts. It is not clear whether these
low- and intermediate-luminosity galaxies contain supermassive black
holes as well. From
an observational perspective, detecting such black holes is extremely
challenging because of their comparatively ``modest'' masses: i.e., for
an intermediate-luminosity galaxy (${\rm M_B} \sim 17.0$ magnitudes
and $\sigma \sim 40$ km s$^{-1}$) 
the M$_{\rm BH}$-$\sigma$ scaling relation predicts M$_{\rm BH} \sim 46,000
- 153,000$ solar masses, with the precise value depending on the choice
of the M$_{\rm BH}$-$\sigma$  relation used for the extrapolation.

Characterizing the central mass concentrations in low- and
intermediate-luminosity galaxies will be a top priority with 30m-class
telescopes, and IRIS will be the instrument of choice for such studies
with TMT. A key open question is the age and abundance distributions
of the nuclei since, in principle, these provide information on the
history of star formation and chemical enrichment on the smallest
spatial scales in galaxies (with obvious implications for the fueling
of the central black hole and AGN feedback). The observations are
challenging, though, because of the small sizes of the nuclei
($0\farcs05 = 4$ pc at the distance of the Virgo Cluster, the nearest
collection of suitable targets), the enhanced ``sky'' brightness due
to the underlying galaxy, and the need for moderate or higher spectral
resolutions ($R \geq 4000 - 8000$).  
The results promise fundamental insights into black hole formation,
however, because different formation scenarios yield very different
predictions for the distribution of black hole masses at the low-mass
end (see Fig.~\ref{fig:BlackHoles}).

\section{Resolved stellar populations in the Local Group} \label{sec:stellarpops}

What are the chemical properties of bulge stars, and how do these vary
with location in a bulge? What fraction of the stars in the inner
regions of bulges originated from material that originated in the
disk, and how far do these stars diffuse from their places of birth?
How do the properties of stars in the Galactic bulge compare with
those in other nearby spheroids? Answers to these questions can be
found by investigating stars in the central regions of
spheroids. Unfortunately, crowding makes it difficult to resolve
individual stars in the central regions of even the closest external
galaxies.  As a result, an understanding of resolved stellar populations
is a major driver on the image quality requirements for IRIS.

With the possible exception of space-based facilities,
efforts to resolve individual stars in the main bodies of bulges are ---
by necessity --- made in the near-infrared wavelength region. Not only
is this the wavelength regime where ground-based telescopes deliver
their best angular resolution, but spheroids tend to be comprised of
stars with relatively high metallicities, and line blanketing can
severely affect the brightnesses of evolved stars with metallicities
that are with a tenth of solar. As a consequence, observations made at
visible wavelength are biased against the detection of the most
metal-rich giants, which will be fainter than their more metal-poor
kin. While the visible wavelength range contains many well-calibrated
features that have traditionally been used for abundance studies, the
K-band also contains lines that can be used to measure the
strengths of various atomic species \cite{Frogel2001}.

Only the very brightest AGB stars can be resolved in the central 100
parsecs of Local Group galaxies with ground-based 4-8 meter
telescopes \cite{Stephens2002, Davidge2000, Davidge2001}. 
Because of the high angular resolution that will be delivered
by TMT+NFIRAOS, IRIS will offer a unique capability for resolving
individual stars in the central regions of M31, M32, and M33, and
characterizing their stellar contents. Not only will the increase in
angular resolution allow deeper observations than those conducted with
4 - 8 meter telescopes, but with integral field spectroscopic 
velocity measurements
one can also identify individual stars in extremely crowded
kinematically hot environments\cite{Davidge2010}.

\section{Kinematics and metallicity variations inside high-redshift galaxies} \label{sec:lbgs}

\begin{figure}
   \begin{center}
   \begin{tabular}{c}
   \includegraphics[width=6.5in]{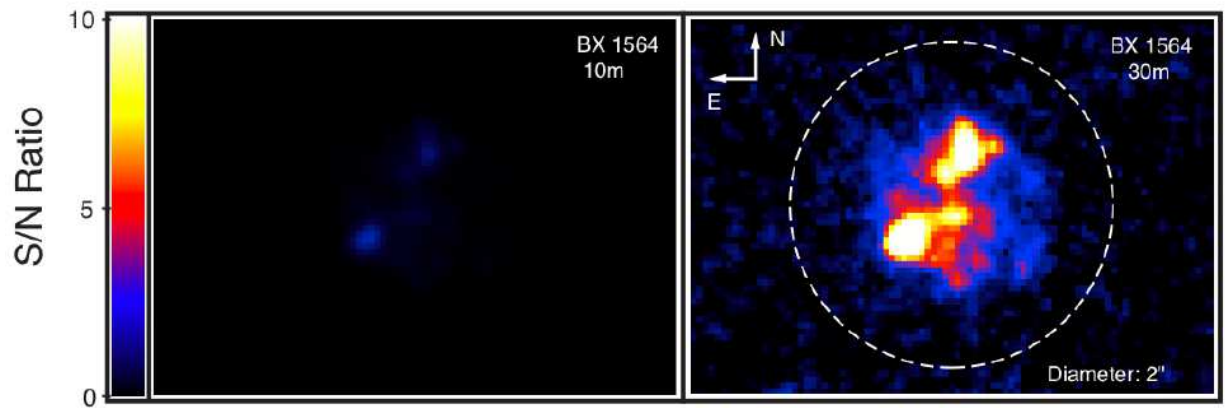}
   \end{tabular}
   \end{center}
   \caption[example] 
   { \label{fig:lbgs} Relative signal-to-noise ratios expected for a
two-hour observation of H$\alpha$ emission from the star-forming galaxy
HDF-BX1564 with current 10m (left-hand panel) and next-generation 30m
(right-hand panel) facilities, from \cite{Law2009b}.  Note the key
increase in the ability to detect low surface brightness emission on
small angular scales between the two galaxy components.}
   \end{figure}

A major breakthrough scientific application of IRIS+NFIRAOS will be
the spatial dissection of galaxies during the peak epoch of galaxy
formation in the range $z \sim 1-5$, the most active known period of
star formation and AGN accretion in the history of the universe.
Observations of these galaxies with the TMT will exploit both its
light gathering power and its unique angular resolution at near-IR
wavelengths.  Large samples of galaxies throughout this redshift range
are already known, and the current generation of 8-10 m telescopes has
recently provided intriguing evidence for prevalent dynamically ``hot''
high-velocity dispersion systems that do not fit neatly into our
current picture of galaxy formation\cite{Law2009a}. TMT and IRIS will allow
observations of a much broader range of systems, and of systems at
higher redshifts (see Fig.~\ref{fig:lbgs})\cite{Wright2010b,Law2009b}.

Spatially resolved spectroscopy of emission lines with the
light-gathering power and spatial resolution of TMT will allow
differences in chemistry, kinematics, and physical conditions to be
mapped as a function of spatial position within the galaxies.  These
measurements are required to go beyond our current knowledge of crude global
properties, gaining a new level of understanding of the physics of
galaxy formation.  

IRIS, with its imaging and IFU capabilities, will provide crucial
steps in a comprehensive survey of these systems.  It will measure
velocity widths and global rotation of galaxies, helping to
distinguish kinematics associated with ongoing merging signatures from
kinematics of rapidly star-forming but otherwise undisturbed galaxies.
By measuring emission line ratios as a function of position within the
galaxies, IRIS will uncover active galactic nuclei that might
otherwise be hidden, and it will probe the spatial evolution of
metallicity gradients within galaxies, which are necessarily closely
tied to their formation mechanisms.\cite{Wright2009,Wright2010a}  The need to sample both the lumpy
and the diffuse parts of high-redshift galaxies with maximal
sensitivity is a significant driver on the IRIS IFU's selectable sampling 
scale, which includes both fine (4, 9 mas) and coarse (25, 50 mas)
sampling.

   \begin{figure}
   \begin{center}
   \begin{tabular}{c}
   \includegraphics[height=7cm]{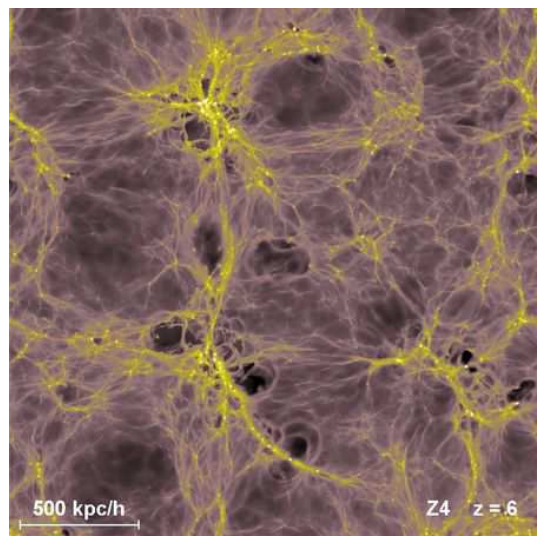}
   \end{tabular}
   \end{center}
   \caption[example] 
   { \label{fig:firstlight} 
A hydrodynamic simulation of the early baryonic density of a $\Lambda$CDM
universe, from \cite{Springel2003}.  At very early epochs, the
star-forming knots in the universe may be detectable as Lyman $\alpha$
emitting sources in the observed-frame near infrared.  IRIS will excel at
detecting and characterizing these sources.}
   \end{figure} 

\section{First light and reionization} \label{sec:firstlight}

Sometime during the first 800 million years after the Big Bang ($z >
7$), the first tiny seeds of galaxies began to collapse and to form
stars.  Understanding the nature of these early galaxies is a
forefront problem in cosmology and astrophysics.  TMT and IRIS will
explore this ``First Light'' epoch by revealing extremely faint
star-forming galaxies and by studying the detailed physical
characteristics of individual early galaxies.

The James Webb Space Telescope (JWST) has the sensitivity and spatial
resolution designed to find early ($z > 7$) galaxies using its
broad-band imaging capabilities.  The capabilities of TMT/IRIS will be
extremely complementary, allowing the discovery of line-emitting
(Lyman $\alpha$) populations that are, most likely, distinct from the
``Lyman break'' galaxy populations JWST will find.  The Lyman $\alpha$
emitters may be younger and lower-mass
objects, potentially containing less metal-enriched and more massive
stars that differ in character from star formation at the current
epoch.  With the most sensitive spectroscopy available in the
near-infrared, IRIS will also study the physical properties of all
types of star-forming first-light galaxies.

The most fundamental question about first light is the number density
of star-forming galaxies at the earliest epochs; these are the key
sources of ionizing photons.  The combination of TMT and IRIS will
contribute most to the problem of finding faint Lyman $\alpha$
emitters in the $7 < z < 19$ range.  By targeting regions around large
galaxies discovered by JWST or the gravitationally lensed regions
around known nearby massive galaxy clusters, we can explore different
regions of the Lyman $\alpha$ luminosity function.

The epoch of the reionization of the universe likely began around the
most rapidly star-forming galaxies and proceeded outward in
``bubbles'' that eventually grew together to fill space. Because Lyman
$\alpha$ emission does not penetrate a neutral medium, the clustering
of Lyman $\alpha$ sources on Mpc scales actually reveals the growth of these
bubbles throughout cosmic time.  Thus, with narrow-band maps of large
areas of sky, IRIS will constrain the luminosity function and
the clustering of Lyman $\alpha$ sources and reveal extensive information
about the nature of reionization.

A very early epoch of reionization of the universe, as indicated by
the results of the WMAP satellite \cite{Spergel2007,Dunkley2009},
requires star formation that is very efficient at emitting ionizing
photons.  In addition, models of the formation of stars in pristine,
metal-free gas suggest that the earliest stars were very massive, and
thus had extremely hard radiation fields.  This ``Population III''
star formation is widely sought as evidence that our theories of the
early epochs in the universe are correct.  Currently, the best
prospect for direct detection of Population III star formation is
spectroscopic evidence from the rest-frame ultraviolet, which has been
redshifted to the near infrared.  Deep IRIS spectroscopy of
pre-selected targets will allow detections of even faint HeII lines
from Population III stars.  Spectroscopic observations of the profile
shape of the Lyman $\alpha$ line will also reveal the characteristics
of the local intergalactic medium.  Taken together, the ``first
light'' science case drives requirements on the sensitivity of IRIS,
as well as the need for high enough spatial resolution to reach
between night sky lines and to resolve the Lyman $\alpha$ line profiles
of faint, high-redshift sources.


\acknowledgments
The authors gratefully acknowledge the support of the
TMT partner institutions. They are the Association of Canadian
Universities for Research in Astronomy (ACURA), the California
Institute of Technology and the University of California.  This work
was supported as well by the Gordon and Betty Moore Foundation, the
Canada Foundation for Innovation, the Ontario Ministry of Research and
Innovation, the National Research Council of Canada, Natural Sciences
and Engineering Research Council of Canada, the British Columbia
Knowledge Development Fund, the Association of Universities for
Research in Astronomy (AURA), the U.S.  National Science Foundation,
and the National Astronomical Observatory of Japan (NAOJ).

\bibliography{IRIS_science}   
\bibliographystyle{spiebib}   

\end{document}